# New Physics in $B_s$ decays to the vector mesons J/ψ and φ

W.N.Cottingham  Physics Department  University of Bristol Tyndall Avenue Bristol BS81TL

Abstract  This paper is concerned with the physics of $B_s$ and $\bar{B}_s$ meson decays to J/ψ and φ. These decays have been observed at the Tevatron [7] and, in particular, should be copiously observed at the LHCb detector. This paper gives a concise presentation of the basic physics analysis and a general formalism for analyzing the signals of CP violation.

The primary object of this paper is to pursue the work of I.Dunietz et al [1] on signals of new physics in $B_s$ decays but gives an extended discussion of J/ψ , φ production with the subsequent electromagnetic and strong decays J/ψ -> $\ell^- \ell^+$ and φ -> $K^-$ $K^+$ . The work of [1] emphasizes the very real possibility of enhanced CP violation in $B_s \bar{B}_s$ mixing. In particular heavy particles, outside of the standard model, could contribute to the box diagrams, and in particular to $M_{12}$ responsible for mixing. With new CP violating phases signals of such new contributions could be observed with the copious event rates in the LHCb detector.  We extend the work of A.Dighe et al [2,3] to give a fuller discussion of the signals of CP violation which could reveal any new physics .We also emphasize the importance of $B_s, \bar{B}_s$ production rate asymmetry which gives access to data otherwise only available through tagging.

The two mass eigenstates of the $B_s, \bar{B}_s$ system are

$|B_L\rangle = p|B_s\rangle + q|\bar{B}_s\rangle$     is the lighter eigenstate

$|B_H\rangle = p|B_s\rangle - q|\bar{B}_s\rangle$     is the heavier

We adopt the convention of Dunietz et al [1], $|\bar{B}_s\rangle = -CP|B_s\rangle$ so that with no CP volation in mixing p = - q

The decay rate of a meson tagged at time t = 0 as a $B_s$ to a final state f is denoted by (See also [1])

$$\frac{d\Gamma}{dt}(B_s, t) = N_f \frac{e^{-\Gamma t}}{4}\left[|A_f + z\bar{A}_f|^2 e^{-y} + |A_f - z\bar{A}_f|^2 e^{+y} + 2\operatorname{Re}((A_f + z\bar{A}_f)(A_f - z\bar{A}_f)^* e^{ix})\right]$$

and for a meson tagged as a $\bar{B}_s$ at time t = 0

$$\frac{d\Gamma}{dt}(\bar{B}_s, t) = N_f \frac{e^{-\Gamma t}}{4|z|^2}\left[|A_f + z\bar{A}_f|^2 e^{-y} + |A_f - z\bar{A}_f|^2 e^{+y} - 2\operatorname{Re}((A_f + z\bar{A}_f)(A_f - z\bar{A}_f)^* e^{ix})\right]$$

$A_f$ and $\bar{A}_f$ are the decay amplitudes for $B_s \to f$ and $\bar{B}_s \to f$ at time $t = 0$, $N_f$ is a normalization factor, $z = q/p$, x and y are time dependent quantities $x = \Delta m_s t$ and $y = \Delta\Gamma_s t/2$, $\Delta\Gamma_s = \Gamma_L - \Gamma_H$

So far experiment is consistent with standard model expectations The D0 collaboration measure $\Delta m_s$ in the interval 17 to 21 ps$^{-1}$ [5]. $\Delta\Gamma_s$ is expected to be in the interval 0.06 to 0.14 ps$^{-1}$ [6] and $\Gamma_s = 0.684 \pm 0.039$ ps$^{-1}$ [4]. At the LHCb for example there is no reason to expect that $B_s$ and $\bar{B}_s$ mesons will be produced in equal numbers, with a significant production rate asymmetry much information can be gleaned without tagging. With a production rate asymmetry Asy, the untagged rate is the sum of the tagged rates with weighting ½ (1 + Asy) and ½ (1 – Asy), all three time dependent terms survive.

The final state f is a charged lepton pair and a K$^-$ K$^+$ pair. We take the J/ψ direction in the Bs rest frame to define the z axis, and the polar angles $\theta_l$ and $\phi_l$ by the negative lepton direction, having moved to the J/ψ rest frame,

$$\hat{p}_{l^-} = -\hat{p}_{l^+} = (\sin\theta_l \cos\phi_l, \sin\theta_l \sin\phi_l, \cos\theta_l)$$

Similarly in the φ rest frame we take the K$^-$ direction to be

$$\hat{p}_{K^-} = -\hat{p}_{K^+} = (\sin\theta_K \cos\phi_K, \sin\theta_K \sin\phi_K, \cos\theta_K)$$

To identify the vector mesons by constructing their masses, these angles, up to an overall azimuthal angle say $\phi_K$, will have to be determined either explicitly or implicitly for every event. We use these angular variables because they explicitly include the important azimuthal angle $\Delta = \phi_l - \phi_K$, to translate to the angles $\theta, \varphi$ and $\psi$ of Dighe et al [1,2,3]

$$\psi = \theta_K, \quad \sin\theta\cos\varphi = \cos\theta_l, \quad \sin\theta\sin\varphi = -\sin\theta_l\cos\Delta, \quad \cos\theta = \sin\theta_l\sin\Delta$$

Neglecting terms of order $\left(\dfrac{m_l}{m_{J/\Psi}}\right)^2$ the helicities of the lepton pair are opposite, the net lepton spin projection m along $\hat{p}_{l^-}$ is either m = +1 or m = -1, we here assume that no lepton helicity is measured and sum these two final spin states.

For m = +1 the decay amplitude for a Bs meson can be taken to be

$$A_f(m=+1) = -(1+c_l)s_K(S-iP)e^{i\Delta} + 2Ls_l c_K + (1-c_l)s_K(S+iP)e^{-i\Delta}$$

S and L are complex parameters that come from even parity (S and D wave) J/ψ φ states, P is a complex parameter from the odd parity P wave state. $c_l$ $s_l$ $c_K$ and $s_K$ are shorthand notations for $\cos\theta_l$, $\sin\theta_l$, $\cos\theta_K$ and $\sin\theta_K$, $\Delta = \phi_l - \phi_K$. The parameters are complex due to phases of the basic production amplitudes but also from the J/ψ and φ propagators and from final state interactions. S – iP, L and S + iP correspond to positive, zero and negative helicity J/ψ φ intermediate states. The CDF collaboration [7] have found that L accounts for 50% or more of the total. For m = -1 we have

$$A_f(m=-1) = -A_f(m=+1, c_l \to -c_l, c_K \to -c_K)$$

Note $s_e$ and $s_K$ are always positive (0 < θ < π). For the $\overline{B}_s$ decay amplitudes

$$\overline{A}_f(m=+1) = -A_f(m=+1, S \to \overline{S}, P \to -\overline{P}, L \to \overline{L})$$
$$\overline{A}_f(m=-1) = -A_f(m=-1, S \to \overline{S}, P \to -\overline{P}, L \to \overline{L})$$

In these last three equations quantities that remain unchanged are not exhibited. With the neglect of direct CP violation $\overline{S} = S, \overline{P} = P$ and $\overline{L} = L$.

We define $S = S_0 + \delta S \quad P = P_0 + \delta P \quad \text{and} \quad L = L_0 + \delta L$
$-z\overline{S} = S_0 - \delta S \quad -z\overline{P} = P_0 - \delta P \quad \text{and} \quad -z\overline{L} = L_0 - \delta L$

CP violation is expected to be small and here we include in the decay rates only terms up to first order in the small quantities $\delta S, \delta P,$ and $\delta L$. The rates are combinations of six angular functions

$$F_1 = \sin^2(\theta_K)(1 - \sin^2(\theta_l)\sin^2(\Delta)) \quad F_2 = \sin^2(\theta_K)(1 - \sin^2(\theta_l)\cos^2(\Delta))$$
$$F_3 = \sin^2(\theta_l)\cos^2(\theta_K) \quad F_4 = \sin^2(\theta_K)\sin^2(\theta_l)\sin(2\Delta)$$
$$F_5 = \sin(2\theta_l)\sin(2\theta_K)\sin(\Delta) \quad F_6 = \sin(2\theta_l)\sin(2\theta_K)\cos(\Delta)$$

Translating these functions to the angles of Dighe et al, they are

$$F_1 = \sin^2(\psi)\sin^2(\theta) \quad F_2 = \sin^2(\psi)(1 - \sin^2(\theta)\sin^2(\varphi))$$
$$F_3 = \cos^2(\psi)(1 - \sin^2(\theta)\cos^2(\varphi) \quad F_4 = -\sin^2(\psi)\sin(\varphi)\sin(2\theta)$$
$$F_5 = \sin(2\psi)\sin(2\theta)\cos(\varphi) \quad F_6 = -\sin(2\psi)\sin^2(\theta)\sin(2\varphi)$$

We find

$$\sum_m |A_f - z\overline{A}_f|^2 = 16\left[2|P_0|^2 F_1 - 2\mathrm{Real}(P_0 \delta S^*) F_4 - \mathrm{Real}(P_0 \delta L^*) F_5\right]$$

$$\sum_m |A_f + z\overline{A}_f|^2 = 16\begin{bmatrix} 2|S_0|^2 F_2 + 2|L_0|^2 F_3 - \mathrm{Real}(L_0 S_0^*) F_6 \\ -2\mathrm{Real}(S_0 \delta P^*) F_4 - \mathrm{Real}(L_0 \delta P^*) F_5 \end{bmatrix}$$

$$2\sum_m (A_f + z\overline{A}_f)(A_f - z\overline{A}_f)^* = 16\begin{bmatrix} -2S_0 P_0^* F_4 - L_0 P_0^* F_5 \\ + 4\delta S^* S_0 F_2 + 4P_0^* \delta P F_1 + 4\delta L^* L_0 F_3 \\ - (\delta S^* L_0 + \delta L^* S_0) F_6 \end{bmatrix}$$

Particularly simple and revealing expressions result if the angles $\theta_l$ and $\theta_K$ and the irrelevant azimuthal rotation about the J/$\psi$ direction are summed over (see also [2]), then

$$\frac{d^2 \Gamma(B_s, t)}{dt \, d\Delta} = N_f' e^{-\Gamma t}\left[A_L(\Delta)e^{-\Delta\Gamma t/2} + A_H(\Delta)e^{+\Delta\Gamma t/2} + A_M(\Delta, \Delta mt)\right]$$

$$\frac{d^2 \Gamma(\overline{B}_s, t)}{dt \, d\Delta} = \frac{N_f' e^{-\Gamma t}}{|z|^2}\left[A_L(\Delta)e^{-\Delta\Gamma t/2} + A_H(\Delta)e^{+\Delta\Gamma t/2} - A_M(\Delta, \Delta mt)\right]$$

with no tagging but with some production rate asymmetry Asy

$$\frac{d^2 \Gamma(B_{untagged}, t)}{dt \, d\Delta} = N_f' e^{-\Gamma t}\begin{bmatrix}(1+\varepsilon - \varepsilon\,\mathrm{Asy})\left[A_L(\Delta)e^{-\Delta\Gamma t/2} + A_H(\Delta)e^{+\Delta\Gamma t/2}\right] \\ +((1+\varepsilon)\mathrm{Asy} - \varepsilon) A_M(\Delta, \Delta mt)\end{bmatrix}$$

$$A_H(\Delta) = |P_0|^2 (2 + \cos(2\Delta)) - 2|\delta S||P_0|\cos(\delta_{P0} - \delta_{\delta S})\sin(2\Delta)$$

$$A_L(\Delta) = |S_0|^2 (2 - \cos(2\Delta)) + |L_0|^2 - 2|\delta P||S_0|\cos(\delta_{S0} - \delta_{\delta P})\sin(2\Delta)$$

$$A_M(\Delta, \Delta mt) = 2 \begin{bmatrix} -|S_0||P_0|\cos(\delta_{P0} - \delta_{S0} - \Delta mt)\sin(2\Delta) \\ |\delta L||L_0|\cos(\delta_{\delta L} - \delta_{L0} - \Delta mt) + |\delta S||S_0|\cos(\delta_{\delta S} - \delta_{S0} - \Delta mt)(2 - \cos(2\Delta)) \\ + |\delta P||P_0|\cos(\delta_{P0} - \delta_{\delta P} - \Delta mt)(2 + \cos(2\Delta)) \end{bmatrix}$$

$\delta_{P0}$ and $\delta_{\delta S}$ are the phases of $P_0$ and $\delta S$ etc $N'_f$ is a normalization factor and $\varepsilon = 1 - \frac{1}{|z|^2}$ is very small.

It is very important to note that in the individual $A_L$ $A_H$ and $A_M$ the CP violating terms have their own characteristic angular dependence which can be used to isolate them. Focusing first on the dominant CP conserving terms, with no tagging but resolving $A_L$ and $A_H$ from the different decay rates $|S_0|^2$, $|P_0|^2$ and $|L_0|^2$ can be determined through the angular dependence. With accurate timing and angular resolution in $\Delta$, $Asy|S_0||P_0|$ can also be measured and the production rate asymmetry inferred.

Returning to the work of I.Dunietz et. al. [1] and to physics beyond the standard model the most likely signal will be in the CP violating terms. In the standard model one can expect only very small direct CP violation The work of [1] and [8] suggests that the most likely signal of new physics will be in the mixing parameter z ( = q/p). With no CP violation in mixing z = -1. The standard model has z very close to -1 and the parameter p = (1 + z)/2 to be almost pure imaginary and of magnitude $2\ 10^{-5}$, plausibly in extended models p could be much enhanced. With no direct CP violation $\delta S/S_0 = \delta P/P_0 = \delta L/L_0 = p$ The work of [1] suggests that p could have a small real part, in their notation and to first order in small quantities

$$p = \frac{1}{2}(i\phi_M + \frac{1}{2}a)$$

This results in the expression

$$\frac{d\Gamma}{dt}(B_s, t) = 4 N_f \, e^{-\Gamma t} \begin{bmatrix} 2 F_1 |P_0|^2 \{e^y - \phi \sin(x-\zeta)\} \\ + 2 F_2 |S_0|^2 \{e^{-y} + \phi \sin(x+\zeta)\} \\ + 2 F_3 |L_0|^2 \{e^{-y} + \phi \sin(x+\zeta)\} \\ + 2 F_4 |S_0 P_0| \{\sin(\delta_1 - x) + \tfrac{1}{2}\phi(e^{-y}\cos(\delta_1 - \zeta) - e^y \cos(\delta_1 + \zeta))\} \\ + F_5 |L_0 P_0| \{\sin(\delta_2 - x) + \tfrac{1}{2}\phi(e^{-y}\cos(\delta_2 - \zeta) - e^y \cos(\delta_2 + \zeta))\} \\ - F_6 |L_0 S_0| \{e^{-y} + \phi \sin(x+\zeta)\} \cos(\delta_2 - \delta_1) \end{bmatrix}$$

$$\phi = \sqrt{\phi_M^2 + a^2/4} \qquad \zeta = \arctan\left(a/2\phi_M\right)$$

$$\delta_1 = \delta_P - \delta_S - \pi/2 \qquad \delta_2 = \delta_P - \delta_L - \pi/2$$

This formula is consistent with that of Dighe et al with
$S_0 = A_\parallel$, $P_0 = -i A_\perp$ and $L_0 = \sqrt{2} A_0$
For a meson tagged as a $\bar{B}_s$ at time t = 0 the time dependent terms involving $x$ ($= \Delta m_s t$) are of opposite sign.

ACKNOWLEDGEMENTS

I thank my colleagues Dr J. Rademacker and Prof. N. Brook for suggesting this work, and Dr A.F. Osorio for helpful correspondence